\documentclass[11pt,a4paper]{article}
\usepackage[margin=2.4cm]{geometry}
\usepackage{amsmath,amssymb}
\usepackage{graphicx}
\usepackage{siunitx}
\DeclareSIUnit{\um}{\ensuremath{\mu}m}
\DeclareSIUnit{\uJ}{\ensuremath{\mu}J}
\usepackage{longtable,booktabs,array,calc}

\providecommand{\real}[1]{#1}
\usepackage{caption}
\usepackage[colorlinks=true,linkcolor=blue,citecolor=blue]{hyperref}

\title{\bf Superradiant temporal transition radiation\\ of a relativistic electron beam}
\author{A.\ Nause$^{1}$ and A.\ Gover$^{2}$\\
\small $^{1}$Physics Department and the Schlesinger Center for Compact Accelerators\\
\small and Radiation Sources, Ariel University, Ariel, Israel\\
\small $^{2}$Department of Electrical Engineering, Physical Electronics, Center for Light-Matter Interaction,\\
\small Tel Aviv University, Israel}
\date{}

\begin{document}
\maketitle

\begin{abstract}
\noindent
A charge in uniform motion carries a bound field that does not radiate. Transition radiation releases part of it at a boundary in space. We propose and analyze the case of radiation emission due to a boundary in time: a relativistic electron bunch passes through an aperture in a semiconductor (silicon) wafer, and while its bound field occupies the wafer, a femtosecond laser switches the semiconductor into an electron-hole plasma. The field is a valid solution of the old medium and not of the new one, so it is handed to the plasma as an initial condition for radiation emission at the plasma frequency as. Because in temporal transition radiation (t-TR) the wavevector is conserved and the frequency is not, the emission can appear in a spectral band that the bunch's own field does not populate. Every electron contributes in phase to the initial condition at the switching moment, so the emitted energy scales as \(N^{2}\). We derive the field energy conversion at the temporal boundary and the resultant measurable amplitude and frequency line-shape of the radiation taking into consideration plasma carriers collisional loss. The simplified analytical model of the process and its scaling laws are confirmed by use of FDTD numerical computations for a reference case based on the parameters of an existing 6.5 MeV photoinjector accelerator. For this case we predict wideband t-TR radiation emission at THz frequencies, where the emission center frequency, near the plasma frequency, is selectable across an octave, set by the laser fluence, and the intensity is governed by the beam charge. The predicted radiative energy is detectable in a single shot. The experimental considerations for observing it are given in a Supplement.
\end{abstract}

\section{Introduction}

A charge moving uniformly in vacuum does not radiate. Radiation requires a broken symmetry: a bent trajectory, a medium in which light is slower than the particle, or, in Transition Radiation (TR), where a material boundary on the path of the charged particle breaks the translational symmetry of space\cite{r1,r2,r3,r4,r5,r6}. Ginzburg emphasized that the last is more general than the foil-crossing picture: transition radiation can take place whenever a source without a proper frequency resides in inhomogeneous or nonstationary media \cite{r7,r8,r9}. The limiting member of that family is a boundary in time. While a spatial boundary conserves frequency and changes wavevector, a temporal boundary conserves wavevector and changes frequency\cite{r10,r11,r12}. At one instant, throughout the region that the field occupies, the medium becomes something else, and the field that was a valid solution a moment earlier would emit radiation at the natural frequencies of what has just been created. Specifically, in the case of a charged particle experiencing a temporal change of the dielectric properties in the region of its bound electromagnetic field, this bound field can turn into radiation. This radiation emission is termed Temporal Transition Radiation (t-TR). In the present paper we suggest a scheme to observe and measure this fundamental radiation phenomenon, enhanced by the process of superradiance\cite{r13}.

Temporal boundaries are now experimentally accessible, in switched microwave metamaterials\cite{r14}, photoinduced interfaces and time-modulated metasurfaces\cite{r15}, and, in prediction, the photonic time crystal, which should amplify the radiation of free electrons traversing it\cite{r16,r17}. In every experiment so far, the input has been a propagating wave. Analysis of radiative emission by a charged particle exists only in theory: for a single slow electron at an optical index step\cite{r18}, and for a bunch at a vacuum-plasma transition with the spatial geometry set aside\cite{r19}. An experimental work related to t-TR is the dc-to-ac converter, in which an ionization front sweeping a biased capacitor array releases its stored field as microwaves\cite{r20,r21,r22}. There the field is a laboratory field, and the boundary is a travelling front. In the present paper the field belongs to the charge and the boundary is temporal: the medium changes at one instant everywhere the field is present, rather than arriving at different places at different times as a front wave does, so the process is a particle process.

\begin{figure}[t]\centering\includegraphics[width=\linewidth]{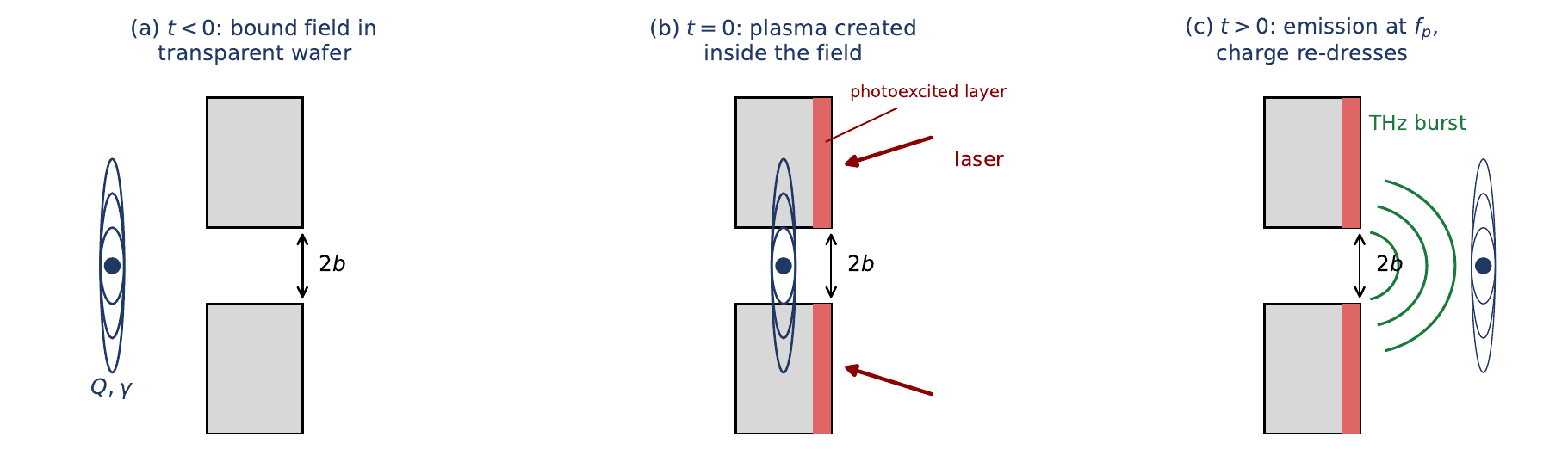}
\caption{Concept. (a) The bound field of an electron bunch occupies a transparent silicon wafer of thickness \(T\) through which the bunch passes via a clearance aperture of radius \(b\). (b) At closest approach an 800 nm laser pulse creates an electron-hole plasma of depth \(d\) in the annulus. The field that was a valid solution an instant earlier is now the initial condition of plasma oscillation in the laser-excited layer. (c) The layer rings at frequency \(\omega_{p}\) and radiates. The bunch has already left; its field re-forms downstream at its own energetic expense.}\label{fig:1}\end{figure}

The configuration of the superradiant t-TR scheme proposed here is shown in Fig.~\ref{fig:1} for a specific example based on the ORGAD RF accelerator \cite{r23}. A relativistic bunch, \(\gamma = 13.7\), 60 pC, 150 fs, passes through a clearance aperture of radius \(b\) in a silicon wafer. Its bound field, Lorentz-flattened to a ``pancake'' of thickness \(b/\gamma\), extends beyond the aperture and fills the wafer. At closest approach an 800 nm pulse switches a layer of depth \(d \simeq 10\si{\um}\) into a plasma of frequency \(f_{p}\) of a few terahertz. The pancake's spectral content lies below \(\gamma c/2\pi b \sim 0.6THz\), so the stored field is nearly at zero frequency relative to the plasma frequency and cannot persist. It is then released radiatively by the plasma at frequency \(f_{p}\).

Two features distinguish this scheme from the single-particle case. The emission is superradiant \cite{r13}: every electron contributes in phase to the initial condition, so the energy scales as \(N^{2}\), here \(N^{2} = 1.4 \times 10^{17}\). But this radiation is superradiant in a different sense from that of bunched beam superradiance\cite{r24,r25}. There the radiation emission is proportional to \(N^{2}\), where \(N\) is the number of electrons in the bunch, and the bunch duration must be shorter than the emitted radiation period, because the bunch is the source. Here the bunch is not the direct source at \(f_{p}\); it sets the initial condition, and the plasma does the emitting. In the present case the \(N^{2}\) dependence on the bunch electrons is valid (even for a bunch longer than the consequent radiation optical period), because what is conserved across the boundary is the stored field, and its energy is proportional to \(N^{2}\). The condition that remains is on the emitting layer, which must be thin compared with the wavelength of the radiation in the medium, so that the charge carriers across the depth of the excited layer will radiate all in phase.

Section~2 gives the mechanism: the bound field and its spectrum, how much of it the dielectric admits, the conversion at the boundary, the competition between radiation and collisions, and the line-shape. In section~3, results of FDTD numerical computation are given, generalizing the analytical results of the model of section 2 and verifying its scaling laws. Section~4 gives the predicted emission for the reference machine. The experimental considerations, backgrounds, discrimination, detection and design choices, are collected in the Supplement.

\section{Mechanism}

\subsection{The bound field and its spectrum}

For a point charge \(q\) moving with \(\beta = v/c\) and \(\gamma = \left( 1 - \beta^{2} \right)^{- 1/2}\), the transverse fields observed at perpendicular distance \(b\) with \(t = 0\) at closest approach, are {[}26, Sec. 11.10{]}

\begin{equation}
E_{\bot}(t) = \frac{q}{4\pi\varepsilon_{0}}\,\frac{\gamma r}{\left( r^{2} + \gamma^{2}v^{2}t^{2} \right)^{3/2}},\quad\quad\mathbf{B} = \frac{\mathbf{v} \times \mathbf{E}}{c^{2}}.
\end{equation}

Its peak and its duration at \(r = b\) are

\begin{equation}
E_{0} = \frac{q\gamma}{4\pi\varepsilon_{0}b^{2}},\quad\quad\tau_{b} \simeq \frac{b}{\gamma c}.
\end{equation}

For a short bunch of \(N\) electrons, each contributes the same field at r = b, so the amplitudes add and \(q \rightarrow Q = Ne\). The condition is that the bunch duration \(\sigma_{t}\) be short against the pancake width, \(\sigma_{t} < \tau_{b} = b/\gamma c\) (not against the radiation emission period as in bunched beam superradiance\cite{r24}). The bunching factor \(\exp\left( - \omega^{2}\sigma_{t}^{2} \right)\) that governs a source emitting at frequency \(\omega\) does not enter here, because the bunch is not radiating at this stage; it supplies a quasi-static field to which all electrons contribute coherently. A bunch comparable to the pancake width reduces the peak field by a computable factor, given in the Supplement, and does not affect the coherence. The plane-wave entry of Sec. 2.2 assumes a wafer thicker than the pancake, so that the whole pancake is inside the dielectric when the switch fires, \(\tau_{b} < T/c\). A membrane thinner than the pancake is treated separately in Sec. 2.2 and Sec. 4.

The finite pancake pulse duration implies a finite bandwidth. Fourier transforming~(1) gives the Weizsäcker-Williams form \(E_{\bot}(\omega) \propto \left( q\omega/\gamma v^{2} \right)K_{1}(\omega b/\gamma v)\) {[}26, Sec. 15.4{]}: a flat spectrum below

\begin{equation}
f_{c} = \frac{\gamma c}{2\pi b}
\end{equation}

and an exponential decay at frequencies above \(f_{c}\),

\begin{equation}
\left| E_{\bot}(\omega) \right|^{2} \propto exp\left( - \frac{2\omega b}{\gamma c} \right).
\end{equation}

Equation~(4) is most significant for the observation and measurement of the t-TR emission process: it guarantees a spectral region in which the bunch itself delivers essentially nothing at the high frequency range through processes like spatial TR that conserve frequency, and it is in this range that the t-TR emission will appear.

\subsection{Field in the dielectric and the stored energy}

For a wafer that is thin compared with the aperture radius, so that matching of the fields at the rim of the annulus can be neglected, but thicker than the pancake, \(T > b/\gamma\), we may use the simplifying model of a plane wave. The more general case of a finite width wafer is treated computationally in Sec. 3. The wafer is perpendicular to the beam and so is the pancake, so the field arrives at the wafer face at normal incidence with \(E_{r}\) and \(H_{\phi}\) both tangential, similarly to a plane-wave pulse. Continuity of both transverse field components forces reflection of the pulse at the wafer face, and the transmitted amplitude is the Fresnel value \cite{r26} \(2E/\left( 1 + \sqrt{\varepsilon_{b}} \right)\). We lump the Fresnel reflection factor with the dielectric constant of the material and present the energy density within the dielectric material as \(u = \frac{1}{2}\varepsilon_{0}\varepsilon_{\rm eff}E^{2}\), where

\begin{equation}
\varepsilon_{\rm eff} = \frac{4\varepsilon_{b}}{\left( 1 + \sqrt{\varepsilon_{b}} \right)^{2}}.
\end{equation}

For silicon \cite{r27}, \(\varepsilon_{b} = 11.7\), then \(\varepsilon_{\rm eff} = 2.4\) and 30 percent of the incident energy is reflected upstream. That reflected pulse is the backward radiation of the wafer in its unswitched state. It is a stationary process, confined by (3,4) to frequencies below the cutoff, \(f_{c}\) so it belongs to the background that the measurement discriminates against and not to the signal.

With \(E(r) = E_{0}(b/r)^{2}\) across the annulus and an excited layer depth \(d\), the charge bound field stored energy just before laser excitation is:

\begin{equation}
U = d\int_{b}^{\infty}\frac{\varepsilon_{0}\varepsilon_{\rm eff}}{2}E_{0}^{2}\frac{b^{4}}{r^{4}}\, 2\pi r\, dr = \frac{\varepsilon_{0}\varepsilon_{\rm eff}}{2}E_{0}^{2}\,\pi b^{2}d\mspace{6mu} \propto \mspace{6mu}\frac{\gamma^{2}Q^{2}d}{b^{2}}.
\end{equation}

In a finite width wafer of width T comparable to \(b\) , the boundary conditions at the aperture rim modify the simplistic plane-wave model, but the scaling law of (6) remains. The numerical solution of Sec. 3 (Supplement, Sec. S9) returns \(\varepsilon_{\rm eff} = 2.3\) to \(2.8\) across the aperture range considered. For a membrane thin against the pulse, the reflections from its two faces partly cancel and the field enters more fully, and consequently \(\varepsilon_{\rm eff} = 7.4\) for a 10 \si{\um} silicon membrane. Eq. 6 indicates that the stored energy is quadratic in charge Q, which carries on to the \(N^{2}\) scaling of the Superradiant energy emitted after the laser excitation; it is inverse quadratic in the aperture radius b, which is the price of clearance; and linear in excited depth \(d\). The excited depth is set by the absorption depth \(1/\alpha\) of the switching pulse. The wafer thickness \(T\) is a separate parameter: it enters in the numerical solution of the field and the computation of the trapped Cherenkov wake (see Supplement). The stored energy does not depend on T once the wafer is thicker than the pancake, \(T > b/\gamma\). For the membrane \(T = d\).

\subsection{Conversion at the temporal boundary}

If the switch is uniform across the illuminated face (transversely) and abrupt in time, \(\mathbf{k}\) is conserved and \(\omega\) is not. A medium that is the same at every point conserves wavevector; a medium that is the same at every time conserves frequency. The reason is that Maxwell's equations, being linear, couple different \(\mathbf{k}\) only through spatial variation of \(\varepsilon\) and different \(\omega\) only through its temporal variation \cite{r19}. Consequently, in t-TR the field keeps its spatial pattern and can exist only at the frequencies that the new dispersion relation assigns to it. In the proposed scheme the temporal transition is assumed to be abrupt, namely shorter than the period of the consequent plasma frequency oscillation. The abrupt change of the medium freezes the field at its instantaneous value and leaves the matching conditions to redistribute it. We present a simplified model analysis for the radiation emission at time t\textgreater0, right after the fast laser switching, assuming the density of the excited layer of width d is uniform. This is not the case in general because the depth profile of the carrier density is graded, \(n(z) \propto e^{- \alpha z}\) (except in the case of a membrane where it is uniform). This may seem to produce spread of wavenumber k and spread of the emitted plasma frequencies along the depth of the layer, which is not the case, as discussed in Sec. 2.5 ahead.

Another point to be considered in the radiative emission model is the spectral contents of the bound field just before switching. As shown above, it has no spectral content above \(f_{c}\) of (3), which is below 1 THz for the apertures considered, while the plasma frequency is several terahertz. In the frequency domain the plasma's permittivity is \(\varepsilon_{b}\left( 1 - \omega_{p}^{2}/\omega^{2} \right)\), negative for every frequency component the field contains, so the low frequency components cannot persist in the interior \cite{r10,r11}, and collateral radiation processes at this frequency band can be ignored. Contrary to the case analyzed in \cite{r18}, the electron beam does not radiate at all after switching and only the carrier plasma oscillates. Based on these considerations and this simplified model, we formulate the oscillation and radiation field development in the excited layer as follows.

Integrating Maxwell's equations across the temporal step gives continuity of the fields \(\mathbf{E}\) and \(\mathbf{B}\), and the carriers are created at rest:

\begin{equation}
\mathbf{E}\left( 0^{-} \right) = \mathbf{E}\left( 0^{+} \right),\quad\quad\mathbf{B}\left( 0^{-} \right) = \mathbf{B}\left( 0^{+} \right),\quad\quad\mathbf{J}\left( 0^{+} \right) = 0.
\end{equation}

The last is the standard suddenly-created-plasma initial state\cite{r28,r29}. For a mode \(\propto e^{ikz}\), Ampère, Faraday and the collisionless equation of motion read

\begin{equation}
\varepsilon_{0}\varepsilon_{b}\dot{E} = - \frac{ik}{\mu_{0}}B - J,\quad\quad\dot{B} = - ikE,\quad\quad\dot{J} = \varepsilon_{0}\left( \omega_{p}^{u} \right)^{2}E,
\end{equation}

with \(\omega_{p}^{u}\) the plasma frequency of the carrier gas before the background permittivity is folded in. To include collisions, we extend the third of these equations according to the Drude equation of motion for the carrier current, \(m^{*}\dot{\mathbf{v}} = e\mathbf{E} - m^{*}\mathbf{v}/\tau\) with \(\mathbf{J} = ne\mathbf{v}\) summed over electrons and holes\cite{r30}. The full form of the dynamic equation is \(\dot{J} + J/\tau = \varepsilon_{0}\left( \omega_{p}^{u} \right)^{2}E\). Here we use the collisionless equation, valid for \(\omega_{p}\tau \gg 1\). The general case is solved numerically in Sec. 3, where \(\omega_{p}\tau\) runs from \(3\) to \(7\) across the tuning range, so the condition \(\omega_{p}\tau \gg 1\) is only marginally satisfied and collisions broaden the line.

Eliminating \(B\) and \(J\) in Eq. 8,

\begin{equation}
\ddot{E} = - \omega_{2}^{2}E,\quad\quad\omega_{2}^{2} = \omega_{p}^{2} + \frac{c^{2}k^{2}}{\varepsilon_{b}},\quad\quad\omega_{p} \equiv \frac{\omega_{p}^{u}}{\sqrt{\varepsilon_{b}}}.
\end{equation}

With the initial conditions (7) and \(B_{0} = \beta E_{0}/c\),

\begin{equation}
E(t) = \frac{E_{0}}{2}\left( 1 - \frac{\beta ck}{\varepsilon_{b}\omega_{2}} \right)e^{i\omega_{2}t} + \frac{E_{0}}{2}\left( 1 + \frac{\beta ck}{\varepsilon_{b}\omega_{2}} \right)e^{- i\omega_{2}t}.
\end{equation}

For a layer thin against \(c/\omega_{p}\) the relevant limit is \(k \rightarrow 0\), where \(\omega_{2} \rightarrow \omega_{p}\) and

\begin{equation}
E(t) = E_{0}\cos\omega_{p}t.
\end{equation}

The entire stored field becomes the amplitude \(E_{0}\) of the oscillation at the plasma frequency: the field cannot jump and the plasma cannot hold a field at zero frequency, so the conversion is of order unity whatever the size of the permittivity change. What the carrier density controls is how fast the field is released. The Maxwell relaxation time of the carrier gas, \(\varepsilon_{0}\varepsilon_{b}/\sigma\) with \(\sigma = \varepsilon_{0}\varepsilon_{b}\omega_{p}^{2}\tau\) the Drude conductivity, is \(1/\omega_{p}^{2}\tau\): 0.34 ps at \(f_{p} = 0.5\ THz\) and 6.5 fs at 3.6 THz. The emission is resonant only when \(\omega_{p}\tau \gtrsim 1\), which sets a floor on the useful carrier density.

\subsection{Radiative escape, collisional loss and the emission lineshape}

Conversion into current oscillation is not the same as conversion into radiation. At \(k\to0\) the oscillation does not propagate. The excited layer is an oscillating current sheet, and its energy leaves either through the two faces of the layer as radiation or into carrier heating through collisions. The rates of the two processes determine whether the layer radiates out and the width of the emission line.

A sheet carrying a transverse surface current K between vacuum and a medium of index \(n_r=\sqrt{\varepsilon_b}\) radiates a transverse field

\begin{equation}
E_{\rm rad}=\frac{Z_0 K}{1+n_r}
\end{equation}

on both sides, since tangential E is continuous through the sheet. This is valid for a large-aperture photoconductive emitter \cite{r31,r32}. The radiative power density is \({E_{rad}}^{2}\)/\(Z_0\) on the vacuum side and \(n_{r}\) \({E_{rad}}^{2}\)/\(Z_0\) on the substrate side, so the energy divides in ratio 1 : \(n_{r}\) and the vacuum side carries a fraction 1/(1 + \(n_{r}\)) (0.23 for silicon). The switched layer is a sheet of this kind, with \(K=Jd\), but unlike the emitters of \cite{r31,r32}, which are driven by a bias field, nothing drives its current. The carriers move only because of the evolving field in the layer, starting from the initial boundary condition at t=0, and this field is also the one the sheet radiates. Relation (12) therefore cannot be read on its own. It evolves with time and has to be solved together with the plasma dynamics.

Two processes run in the layer simultaneously. The carriers accelerate in the field, following the Drude equation of (8) with its collision term restored,

\begin{equation}
\frac{dJ}{dt}=\varepsilon_0\varepsilon_b\omega_p^{2}E-\frac{J}{\tau},
\end{equation}

and the field drains out through the faces, following Ampère's law integrated across the layer with outgoing waves on both sides,

\begin{equation}
\varepsilon_0(\varepsilon_b-1)\,d\,\frac{dE}{dt}=-J\,d-\frac{(1+n_r)E}{Z_0}.
\end{equation}

The left-hand side of (14) is the displacement current of the material polarization. The first term on the right is the carrier current and the second is the field leakage through the faces, which is (12) turned around: the two half-spaces draw a surface current \((1+n_r)E/Z_0\) from whatever field sits at the sheet. The vacuum part of the displacement current is the forward outgoing wave itself and is not stored in the layer, which is why \(\varepsilon_{b}\) - 1 appears on the left hand side of (14) rather than \(\varepsilon_{b}\). Eliminating J gives a single equation for the field,

\begin{equation}
\begin{gathered}\frac{d^{2}E}{dt^{2}}+2\Gamma\frac{dE}{dt}+\omega_0^{2}E=0,\\ \omega_0^{2}=\frac{\varepsilon_b}{\varepsilon_b-1}\,\omega_p^{2}+\frac{\gamma_r}{\tau},\qquad \gamma_r=\frac{(1+n_r)\,c}{(\varepsilon_b-1)\,d},\qquad \Gamma=\frac{\gamma_r+1/\tau}{2}.\end{gathered}
\end{equation}

This is a damped oscillator equation with two damping rates, the radiation escape rate \(\gamma_{r}\) and the collision loss rate \(1/\tau\), and a natural frequency \(\omega_{0}\) close to the plasma frequency \(\omega_{p} = {2\pi f}_{p}\). \(\Gamma\) is half the total damping rate. The solution of the second-order equation (15) is \(E(t) = e^{- \Gamma t}(A\cos\omega_d t+B\sin\omega_d t)\) with \(\omega_{d} = \sqrt{\omega_0^{2}-\Gamma^{2}}\). The collision rate \(1/\tau\) is the same at every depth. The radiative rate \(\gamma_{r}\) is the rate at which the polarization charge per unit area of the layer, \(\varepsilon_0(\varepsilon_b-1)dE\), would discharge through the leakage current \((1+n_r)E/Z_0\) of the two half-spaces if the carrier current were zero. Since the leakage is set by E alone while the stored charge grows with d, \(\gamma_{r}\) falls with depth.

The quality factor of the oscillator of (15) is the oscillation frequency divided by the energy decay rate,

\begin{equation}
Q=\frac{\omega_p}{\gamma_r+1/\tau},\qquad \frac{1}{Q}=\frac{1}{Q_{\rm rad}}+\frac{1}{\omega_p\tau},
\end{equation}

where

\begin{equation}
Q_{\rm rad}\equiv\frac{\omega_p}{\gamma_r}=\frac{(\varepsilon_b-1)\,\omega_p d}{c\,(1+n_r)}.
\end{equation}

is the quality factor in the absence of collisions: the number of radians (\(Q_{\rm rad}/2\pi\) cycles) through which the field oscillates during the radiative energy decay time \(1/\gamma_{r}\). The quality factor scales in proportion to the layer depth d and to the plasma frequency \(\omega_{p}\). For the thick wafer case with d = 10 \si{\um} on the silicon face, \(n_r=3.42\), the radiation escape time is \(1/\gamma_{r}\) = 81 fs independent of frequency, and \(Q_{\rm rad}\) = 1.8 at \(f_{p}\) = 3.6 THz and 0.86 at \(f_{p}\) = 1.7 THz. The condition for an observable peak in the radiation spectrum is Q \textgreater{} 1/2. Even without collisions, therefore, the thick wafer case would display only a marginally resonant spectrum at the low end of the tuning frequency range. On the other hand, in the membrane case, T = d = 10 \si{\um}, there is an enhancement of \(Q_{\rm rad}\) by a factor 2.21, simply because there is vacuum on both sides of the wafer and \(n_{r}\) = 1 in (17). Consequently \(Q_{\rm rad}\) = 4.0 at 3.6 THz and 1.9 at 1.7 THz for the same frequencies. In Secs. 3 and 4 we consider the spectral observability for this preferable option of a membrane.

Collisions add a term \(1/\tau\) to the radiative rate \(\gamma_{r}\) in (15). With \(\tau=0.3\)~ps for the Drude collision time of carriers in silicon at terahertz frequencies \cite{r27,r33}, the overall quality factor Q of (16) deteriorates from \(Q_{\rm rad}\) to \(Q=0.7\) at 1.7 THz and 1.4 at 3.6 THz for the thick wafer, and to 1.2 and 2.5 for the membrane. Collisions also divert part of the stored energy to heat. In the resonant regime the fraction that leaves as radiation is the ratio of the two rates,

\begin{equation}
f_{\rm rad}=\frac{\gamma_r}{\gamma_r+1/\tau}=\frac{\omega_p\tau}{\omega_p\tau+Q_{\rm rad}},
\end{equation}

The fraction is largest near critical damping, \(Q_{\rm rad}\) = 1/2, where the field leaves the layer fastest relative to the collision time, and falls on both sides of it: a thinner layer decays slowly and loses to collisions, a deeper one oscillates longer and loses to collisions. The depth and fluence dependence of Q and \(f_{\rm rad}\) is tabulated in the Supplement.

There is another condition that caps the depth of the plasma layer d from above. For the entire depth of the layer to radiate in phase, the propagation phase across the layer, \(2\pi d/\)\(\lambda_{m}\) with \(\lambda_{m} = c/(f_p\sqrt{\varepsilon_b})\) the wavelength in silicon, must stay below \(\pi \), so d \textless{} \(\lambda_{m}/2\), which is 12 \si{\um} at 3.6 THz. Between the critical depth at which \(Q_{\rm rad}\) = 1/2, 2.7 \si{\um} at 3.6 THz, and \(\lambda_{m}/2\) the layer radiates as one surface current sheet. The optimal choice of a 10 \si{\um} layer in the present design example sits inside that window across the terahertz range.

The choice of a d = 10 \si{\um} layer in the present example therefore corresponds to the resonant regime, but only marginally. With collisions, (16) gives \(Q=0.7\) at 1.7 THz and 1.4 at 3.6 THz, so the amplitude falls by \(1/e\) within 0.2 to 0.5 of a cycle. What is emitted is a single damped cycle rather than a train. The spectral line is correspondingly wide, \(\Delta f/f\approx1/Q\), of order one, with a resolvable peak whose center is the observable. The lineshape itself is computed and reported in Sec. 3.

\subsection{Emission frequency}

The photoexcited electron-hole gas responds as a Drude plasma at terahertz frequencies \cite{r27,r33,r34}, with both carriers contributing,

\begin{equation}
\omega_p^{2}=\frac{ne^{2}}{\varepsilon_0\varepsilon_b}\left(\frac{1}{m_e^{*}}+\frac{1}{m_h^{*}}\right),
\end{equation}

with the conductivity effective masses \({m_{e}}^{*}\) = 0.26 \(m_{e}\) and \({m_{h}}^{*}\) = 0.37 \(m_{e}\) of silicon \cite{r35}. This is the \(\omega_{p}\) of (15) to (17). The switching pulse has fluence F, the energy per unit area, and photon energy \(E_{\rm ph}\). With one electron-hole pair generation per absorbed photon \cite{r32}, the carrier density that the laser beam creates at the surface is

\begin{equation}
n_0=\frac{F(1-R)\,\alpha}{E_{\rm ph}},
\end{equation}

with R the surface reflectivity and \(\alpha \) the absorption coefficient at the pump wavelength. In silicon at 800 nm, \(R=0.33\), \(\alpha^{-1}=12\)~\si{\um} \cite{r36} and \(E_{\rm ph}\) \(E_{\rm ph}=1.55\)~eV, so a fluence of 100 \si{\uJ\per cm^2} creates \(n_0\) = 2.3 \(\times \) \(10^{17}\) \(cm^{-3}\) at the surface and a surface plasma frequency \(f_{p0}\) = 3.2 THz. The emission frequency is therefore set by the laser, through \(\omega_{p}\) \(\propto \) \(\sqrt F\), and is independent of the electron beam.

A note is in order on the effect of the longitudinal density variation on the emission frequency. Because n falls as \(e^{- \alpha z}\) across the excited layer, the local plasma frequency is \(\omega_{p0}e^{- \alpha z/2}\), and treating the oscillating current as an assembly of transverse slices, each oscillating at its own frequency, would predict a spectrum spread from \(\omega_{p0}\) downward. That would be the case if the slices were independent oscillators, and they are not. The layer is thinner than half the wavelength in the medium, which is the condition of Sec. 2.4 that is required for the entire depth of the layer to radiate in phase, so the slices share a single field and oscillate together at the plasma frequency of their mean density. For the 10 \si{\um} layer, \(\alpha d=0.85\), the mean density is 0.67 of the surface value, which places the spectral line near \(\sqrt{0.67}\) = 0.82 \(f_{p0}\). The spectral line center frequency is therefore determined by the laser fluence, a fixed fraction below the plasma frequency at the surface of the wafer.

The spectral width of the line comes from the two rates of Sec. 2.4 and not from the density gradient. Since \(\gamma_{r}\) does not depend on \(\omega_{p}\) and \(1/\tau\) is fixed, the absolute spectral width is constant across the tuning range, Q grows in proportion to the laser-fluence-controlled frequency, and the line sharpens as it tunes up.

\section{Numerical solution}

The analytic model treats the bunch as a point particle and the wafer as a face without a rim. Both approximations are removed by solving Maxwell's equations directly. We use a two-dimensional cylindrical finite-difference time-domain (FDTD) scheme in the scattered-field formulation \cite{r37,r38}: the incident field is known in closed form at every point, and the grid carries only the increment between the total field and the incident field. This kind of formulation is commonly used for computation of beam-driven wakefields \cite{r39}. In our computation model, a 3-D field distribution signal (the ``pancake'' of Fig.1) is incident and scattered by the finite width apertured wafer. Its profile is the analytic field (1) of the macro-charge, convolved along \(z\) with the Gaussian bunch profile, and this field propagates in time and space with a rigid z \(-\) vt spatio-temporal dependence, so the convolution can be done in z or t. The polarization of the wafer material, \(\left( \varepsilon_{b} - 1 \right)\varepsilon_{0}\mathbf{E}\), drives the scattered field. No moving charge is placed on the grid, which avoids the numerical Cherenkov instability \cite{r40,r41}, the spurious growth that arises when a charge advanced on a grid couples to grid modes whose numerical phase velocity is below the particle velocity, while retaining the full geometry and the trapped Cherenkov wake that a thick wafer supports: inside silicon the bunch field propagates at c/\(n_r\), slower than the bunch, so the wafer carries a Cherenkov-like wake at 73° to the axis which is totally internally reflected and stays in the wafer (Supplement). The field energy in the photoexcited layer at the illuminated layer as a function of time just before the illumination is

\begin{equation}
U(t) = \int_{layer}^{}\frac{1}{2}\varepsilon_{0}\varepsilon_{b}\,\left| \mathbf{E}(r,z,t) \right|^{2}\, 2\pi r\, dr\, dz,
\end{equation}

The peak of this energy at t=0 is the energy handed to the plasma if the switch is timed at that instant. The entry factor (\(\varepsilon \)\_eff = 2.3 to 2.8 against the Fresnel 2.4), the aperture scaling of the width and peak of U(t), and the bunch-length factors that the energies of Table 1 include are given in the Supplement, Sec. S9, together with the validation of the numerical solution against the analytic model, mesh refinement, scale invariance and wafer thickness \(T\).

\begin{figure}[t]\centering\includegraphics[width=\linewidth]{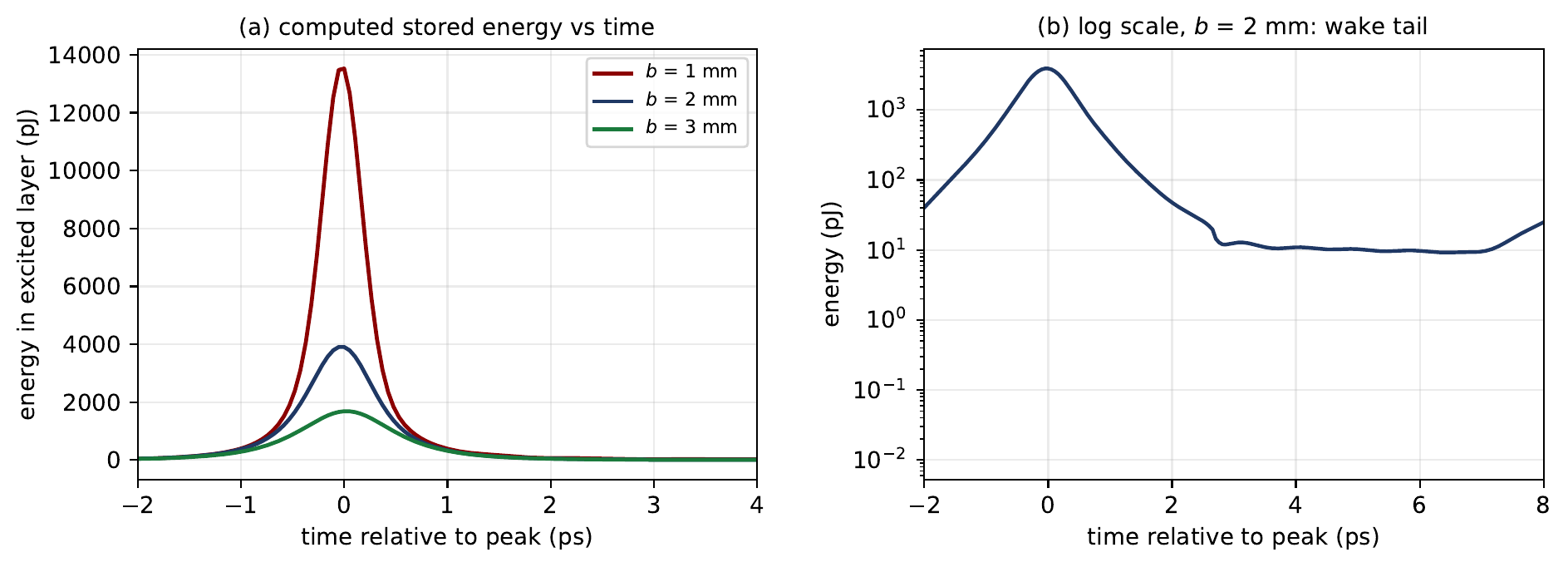}
\caption{Energy in the photoexcited layer (before excitation) against time for the 150 fs bunch, a 500 \si{\um} wafer with a 10 \si{\um} excited layer, computed with a 25 \si{\um} mesh, for three aperture radii. (a)The stored energy \(U(t)\), integrated over the excited layer; the peak value is the energy available to the plasma, and the width is the range of switch delays over which that energy can be caught. (b) The same for \(b = 2mm\) on a logarithmic scale, showing the trapped Cherenkov wake as a plateau at 0.2\% of the peak.}\label{fig:2}\end{figure}

Fig.~\ref{fig:2} presents the result of the FDTD numerical computation for the reference machine, the ORGAD hybrid photoinjector at Ariel University \cite{r23}, whose beam parameters are E=6.5 MeV, Q=60 pC and \(\sigma_{t} =\)150 fs, passing a silicon wafer of thickness T=500 \si{\um} with a d=10 \si{\um} excited layer. It displays the stored energy \(U(t)\) integrated over the excited layer for three values of the aperture radius (b=1mm, 2mm and 3mm) and the same on a logarithmic scale (for b=2mm).

The plasma dynamics and the radiation after the switch, t \textgreater{} 0, are computed by a one-dimensional solution of Maxwell's equations with the Drude current (13) along the depth of the layer, with the field distribution at t = 0 taken from the pre-switching two-dimensional run. For the Drude collision time of silicon, \(\tau \) = 0.3 ps \cite{r33}, it gives directly the radiated fraction, the spectral line centre and the linewidth; the results are collected in Tables~\ref{tab:1} and~\ref{tab:2} of Sec.~4. The radiated fraction used there is that of the stored energy alone, Eq. (18); the solver's total, which also counts the reflection of the part of the pulse still arriving after the switch, is higher (Supplement, Sec. S9).

\section{Predicted emission}

For the reference machine, the ORGAD hybrid photoinjector at Ariel University \cite{r23}, a 6.5 MeV, 60 pC, 150 fs beam and a 10 \si{\um} excited layer in silicon, the predicted emission is collected in Table~\ref{tab:1}. The stored energy at the switch instant, from the FDTD solution of Sec. 3 (Fig.~\ref{fig:2}), runs from 22.4 nJ at b = 0.75 mm to 1.8 nJ at b = 3 mm. Of this a fraction \(f_{\rm rad}\) = 0.79 radiates, 0.23 of it into the vacuum half-space for a thick wafer, and a collection and transmission efficiency of 0.08 (Supplement) brings 25 to 319 pJ to the detector. A pyroelectric detector's single-shot detection threshold is of order 1 pJ, so the signal is detectable in a single shot at every aperture, with a margin of one to two and a half orders of magnitude.

\begin{table}[t]\centering\small
\caption{Predicted yield for the reference beam and a 10~\si{\um} excited layer, from the two-dimensional solution. \(U\) is the stored energy at the switch instant, \(f_{\rm rad}\) the radiated fraction of (18), and the forward share is the vacuum-side fraction \(1/(1+n_r)\). The detected energy estimate is based on a collection and transmission efficiency of 0.08. The membrane was computed at \(b=1\)~mm only; its aperture dependence follows the same scaling as the wafer, since the entry factor does not depend on \(b\).}\label{tab:1}
\begin{tabular}{lccccc}\hline
target & \(b\) (mm) & \(U\) (nJ) & \(f_{\rm rad}\) & forward share & detected (pJ)\\ \hline
Si wafer, 500~\si{\um} & 0.75 & 22.4 & 0.79 & 0.23 & 319\\
 & 1.0 & 14.5 & 0.79 & 0.23 & 207\\
 & 1.5 & 7.2 & 0.79 & 0.23 & 103\\
 & 2.0 & 4.2 & 0.79 & 0.23 & 59\\
 & 3.0 & 1.8 & 0.79 & 0.23 & 25\\
Si membrane, 10~\si{\um} & 1.0 & 42 & 0.65 & 0.5 & 1100\\ \hline
\end{tabular}\end{table}

\begin{figure}[t]\centering\includegraphics[width=\linewidth]{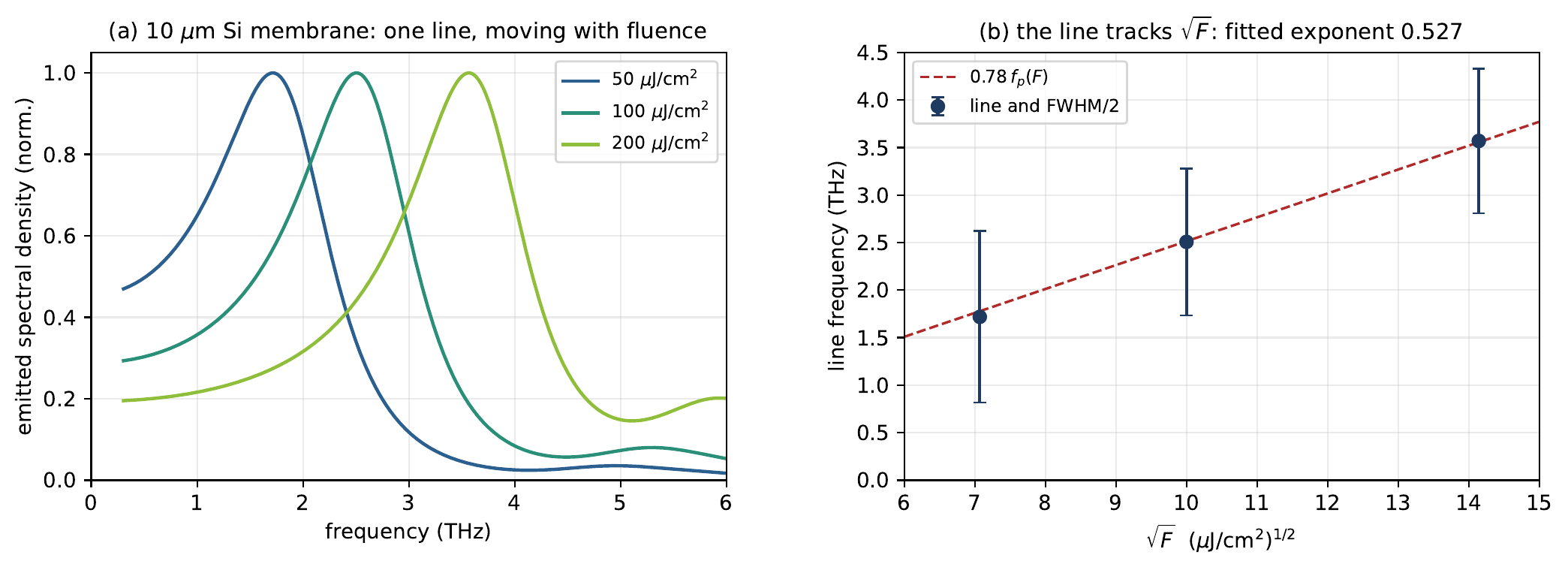}
\caption{Predicted spectroscopy with a free-standing 10 \si{\um} silicon membrane switched by a 800 nm ultrafast laser pulse, as predicted from the one-dimensional switched-Drude solution. (a) Emitted spectral density, normalized, for three fluence values: the spectral peak slides from \(1.7\) to 3.6 THz the higher is the fluence. (b) Computed line frequency center against \(\sqrt{F}\) . The three computed data points follow a proportionality relation \(0.78\, f_{p0}\) and a power-law fit f \(\propto \) F\^{}x to the three points gives x = \(0.53\) (0.5 expected); bars mark the half width at half maximum of the peak.}\label{fig:3}\end{figure}

A free-standing membrane does better than the thick wafer for three reasons. Both faces face vacuum, so half of the emitted t-TR radiation goes forward instead of 0.23 in a thick silicon wafer. The field gets in more fully: \(\varepsilon_{\rm eff}\) = 7.4 instead of 2.4. These two factors enhance the field proportionally. A third advantage is enhancement of \(Q_{\rm rad}\) : Substitution of \(n_{r} = 1\) in Eq. 17 instead of 3.42 enhances \(Q_{\rm rad}\) by a factor of 2.2. This provides longer oscillation time and sharper spectral resonance, but at the same time slightly reduces the escaped radiation amplitude because collisions take a bigger fraction of the energy (Eq. 18). For the prescribed parameters and \(\tau = 0.3\ ps\), the radiation escape factor \(f_{\rm rad}\) falls from 0.79 to 0.65.

For an aperture radius b = 1 mm, the d = 10 \si{\um} membrane stores 42 nJ. It delivers 1.1 nJ to the detector, five times what the thick wafer gives at the same aperture. Its spectrum has a resolved peak (Fig.~\ref{fig:3}a, Table~\ref{tab:2}), with Q = 1.0 at 50 \si{\uJ\per cm^2} rising to 2.4 at 200 \si{\uJ\per cm^2}: a damped transient of one to two cycles. Its spectral line center moves in proportion to the square root of the fluence F over an octave: 1.7 THz at 50 \si{\uJ\per cm^2}, 3.6 THz at 200 \si{\uJ\per cm^2} (Fig. 3b). The line sits at 0.78 \(f_{p0}\) of the surface plasma frequency of Sec. 2.5, and a power-law fit \(f \propto F^{x}\) to the three computed points gives x = 0.53, against the expected 0.5. Across this range the radiated fraction stays at \(f_{\rm rad}\) = 0.65, since by (18) it does not depend on the fluence. Q falls under 1/2 below about 8 \si{\uJ\per cm^2}, and the peak is resolved, Q \(\gtrsim \) 1, from 50 \si{\uJ\per cm^2} upward, so the octave above it is usable. For the thick wafer case both fluences are about three times higher. The spectrum is measured with a bandpass filter set in front of the pyroelectric detector, and the field waveform, from which the lineshape follows, by electro-optic sampling in GaP with a second pickoff of the laser (Supplement).

\begin{table}[t]\centering\small
\caption{The membrane spectral line emission parameters against the laser switching fluence, from the one-dimensional solution (Fig.~\ref{fig:3}). \(f_{p0}\) is the surface plasma frequency from (19) and (20); \(Q\) is the line centre divided by its full width at half maximum.}\label{tab:2}
\begin{tabular}{cccccc}\hline
\(F\) (\si{\uJ\per cm^2}) & \(f_{p0}\) (THz) & line centre (THz) & FWHM (THz) & \(Q\) & \(f_{\rm rad}\)\\ \hline
50 & 2.28 & 1.72 & 1.8 & 1.0 & 0.65\\
100 & 3.22 & 2.52 & 1.56 & 1.6 & 0.65\\
200 & 4.56 & 3.57 & 1.5 & 2.4 & 0.65\\ \hline
\end{tabular}\end{table}

This line frequency dependence on the laser fluence is a signature of the t-TR radiation and would help in observing and measuring the effect in distinction from other present radiation mechanisms. In our scheme, the laser sets the frequency and the electron beam sets the amplitude. Also, it was pointed out that the frequency band of the t-TR radiation is one in which the bunch's own field is suppressed by the decay factor (4), by ten orders of magnitude (at 3 THz and \(b = 2mm\)), and therefore radiation mechanisms associated with the electron beam, like spatial TR, do not overcast the predicted t-TR spectrum. Diffraction radiation from the wafer, the Cherenkov wake it traps, and the laser's own terahertz emission are each either stationary, and so confined below \(f_{c}\), or independent of the beam, and so removed by the laser-only shots interleaved with the coincidences. The separation of the signal from each of these, the collection optics for the hollow emission cone, the detection chain, which is that of the machine's terahertz FEL programme\cite{r42,r43,r44}, and the choice of wafer, membrane and pump wavelength are set out in the Supplement, together with the design consequences of the ceiling \(d < \lambda_{m}/2\) and the floor \(\omega_{p}\tau \gtrsim 1\) derived above.

\section{Conclusion}

Transition radiation at a boundary in time (t-TR) has been analyzed in this article for a case that has not been treated so far: a relativistic electron bunch releases its bound field coherently when the dielectric medium (a semiconductor slab) holding it is switched into an electron-hole plasma. The conversion efficiency of the field energy bound to an electron beam bunch into t-TR radiation is of order unity, because the bound field has no spectral content near the terahertz plasma frequency: it cannot jump across the abrupt temporal boundary, and the plasma cannot hold it. The emission is Superradiant and proportional to the number of electrons squared but the bunch is not bound to be shorter than the emitted radiation period (as is the case in bunched beam superradiance \cite{r24}), because the bunch field is the initial condition for the t-TR process in the plasma medium and not the direct source of radiation. In the t-TR radiation emission scheme proposed in this article, the emission frequency is set by the carrier density of electron-hole plasma in a semiconductor wafer excited by an ultra-short laser pulse, and so it is governed by the laser intensity; the amplitude of the emitted radiation is set by the electron beam. For a reference case based on an existing 6.5 MeV photoinjector, the predicted t-TR radiation in the presented scheme is a terahertz transient of one to two cycles, tunable across an octave at fixed beam, in a spectral band where no stationary process can follow it and obscure the measurement. The predicted radiative energies in this scheme are high enough to be detectable in the lab in a single shot, and the spectrum can be resolved with bandpass filters or by electro-optic sampling. No such t-TR emission process has been observed for a free charge in any regime.

Acknowledgement

We acknowledge the support of the Israel Science Foundation (Grant No. 1705/22).

\clearpage
\setcounter{figure}{0}\setcounter{table}{0}\setcounter{equation}{0}
\renewcommand{\thefigure}{S\arabic{figure}}
\renewcommand{\thetable}{S\arabic{table}}
\renewcommand{\theequation}{S\arabic{equation}}
\begin{center}{\Large\bfseries Supplementary Material}\\[4pt]
{\large Considerations for experimental demonstration and measurement}\end{center}
\medskip
\noindent This Supplement collects what concerns the demonstration and measurement of the effect described in the main text. Section and equation numbers without the prefix S refer to the main text. References [1] to [44] are those of the main text; references [45] to [47] follow them in the common list.

\section*{S1. Approximations of the bunch field}

The analytic estimates take the bunch field to be the transverse field of a line charge on the axis. Three things are left out. Two are negligible and one changes the numbers.

The axial component, \(E_{z} = q\gamma z\prime/4\pi\varepsilon_{0}\left( b^{2} + \gamma^{2}z\prime^{2} \right)^{3/2}\) with \(z\prime = z - vt\), gives \(E_{z}/E_{\bot} = z\prime/b\). It is antisymmetric about closest approach, peaks at \(0.385/\gamma = 2.8\) per cent of the transverse peak, and carries \(1/3\gamma^{2} = 0.18\) per cent of the transverse field's energy. It vanishes in the wafer plane at the instant the switch fires, and what little enters the layer drives a longitudinal current that excites the bulk plasmon, which does not radiate from a planar sheet. The two-dimensional solution carries it exactly; the analytic estimates drop it.

The transverse size of the bunch needs no correction. Averaging the rim field over a Gaussian of rms radius \(\sigma_{r}\) gives \(1 + \frac{3}{2}\left( \sigma_{r}/b \right)^{2}\) times the point value, positive because the near side of the distribution gains more than the far side loses, so the line-charge estimate is conservative. The energy is three per cent high at \(\sigma_{r} = b/10\) and sixteen at \(b/5\). Charge that reaches the rim strikes the wafer and is the halo of Sec.~S3, not a field correction.

The bunch length does change the numbers. The pancake at radius \(b\) has a full width of \(1.53\, b/\gamma c\), the bunch a full width of \(2.35\,\sigma_{t} = 0.35ps\), or 106 \(\mu m\). At \(b = 1mm\) the two are equal and the field of the bunch, the point-charge field convolved along \(z\) with the bunch profile, peaks at \(0.72\) of the point value. The stored energy falls less, to \(0.69\), because the outer radii of the annulus have wider pancakes and are less affected. The factors \(\eta\) in Table~S2, \(0.60\) to \(0.93\) from \(b = 0.75\) to 3 mm, come from the two-dimensional solution on a 25 \(\mu m\) mesh and agree with a direct convolution to two per cent. The membrane's factor at \(b = 1mm\) is \(0.76\), larger than the wafer's, because the switch catches the pancake at mid-plane.

\section*{S2. The Cherenkov wake}

In silicon \(\beta n = 3.41 > 1\), so the bunch is superluminal in the medium and drives a Cherenkov wake at \(\theta_{C} = \arccos(1/\beta n) = 73^{\circ}\), the regime of the dielectric wakefield accelerator\cite{r45,r46}. Inside the wafer the field is propagating rather than evanescent, so the aperture cutoff (4) does not apply to it, and whatever wake is present at the switch instant is converted alongside the quasi-static field. The analytic stored energy (6) omits it and is a lower bound; the two-dimensional solution includes it, and it is the excess above the Fresnel value at the smallest apertures in Sec.~2.2.

The wake does not threaten the measurement, for two independent reasons. It is a stationary process, driven by the beam's own spectral content, so it carries the bunching factor \(\exp\left( - \omega^{2}\sigma_{t}^{2} \right)\) and dies above \(1/2\pi\sigma_{t} = 1.06THz\), a factor \(3.4 \times 10^{- 4}\) at 3 THz. And it propagates at 1270 mrad from the axis, two orders of magnitude outside the forward collection cone of Sec.~S3.

What it does add is a residual in one null test. At \(73^{\circ}\) the wake exceeds silicon's critical angle of \(17^{\circ}\), is totally internally reflected, and escapes at the wafer edge after of order 100 ps. The delay scan therefore has a weak tail behind the prompt peak, and the switch-late null returns a small residual rather than zero. The two-dimensional solution puts it at \(0.2\) per cent of the prompt peak for \(T = 500\mu m\). A thickness series from 100 to 500 \(\mu m\) separates prompt from trapped contributions. The 10 \(\mu m\) membrane is thinner than a terahertz wavelength, traps nothing, and its switch-late null is clean.

\section*{S3. Realisation on the ORGAD photoinjector}

The reference machine is the ORGAD hybrid S-band photoinjector\cite{r23}, with the quasi-optical transport, detectors and diagnostics of its terahertz FEL programme\cite{r42,r43,r44}. The beam is 6.5 MeV (\(\gamma = 13.7\)), 60 pC (\(N = 3.75 \times 10^{8}\)), \(\sigma_{t} \simeq 150fs\), at 1 Hz. It passes through a clearance aperture of radius \(b\) in the target, so no charge intercepts material. Two targets of one material serve two purposes: a thick silicon wafer, excited over a 10 \(\mu m\) layer, the absorption depth at 800 nm being 12 \(\mu m\), as the absolute energy reference, and a free-standing 10 \(\mu m\) silicon membrane, both faces in vacuum, for the spectroscopy and the highest yield.

The switch is a pickoff of the photocathode drive laser, an 800 nm Ti:sapphire system delivering 6 mJ in 35 fs, measured by autocorrelation\cite{r47}, at 500 Hz, whose third harmonic drives the cathode. The pickoff is expanded over about 1 \(cm^{2}\) and attenuated to 10 to 200 \(\mu J/cm^{2}\). Because the same laser creates the bunch, the switch is synchronised to it at the femtosecond level by construction. A delay stage sets the switch time to \(\pm 5ps\) in 50 fs steps. The 35 fs pulse is short against both the passage time, 0.49 ps at \(b = 2mm\), and the plasma period, about 300 fs at 3 THz, so the step approximation of (7) holds with an order of magnitude to spare. The switch must also be simultaneous across the face: at \(20^{\circ}\) from normal the arrival time varies by 4.6 ps across a 4 mm annulus, which would turn the temporal boundary into a travelling ionisation front. Near-normal incidence or pulse-front-tilt compensation is required.

The emission is a hollow cone. The excited annulus carries a radial current, a ring of outward dipoles whose opposite sides cancel on axis, and the intensity rises off axis to a peak near \(\lambda/2\pi b\): 26 mrad for the 1.7 THz line and 12 mrad for 3.6 THz at \(b = 1mm\), with ninety per cent of the power inside 120 and 63 mrad. The cone resembles ordinary transition radiation but its angle is set by diffraction at the aperture and moves as the line tunes, whereas a foil's is fixed at \(1/\gamma = 73mrad\). An off-axis parabola at \(f/2\), accepting \(\pm 250mrad\), collects essentially all of it at every frequency, so the energy measurement does not depend on where the line sits. The null on axis is the radial polarisation made visible. For the membrane the forward and backward cones carry half each.

Detection uses three instruments on the collected beam. A calibrated pyroelectric detector behind a bandpass filter set is the energy channel, single-shot at its picojoule threshold and gateable at the laser rate. A liquid-helium bolometer, NEP of order \(10^{- 13}\) W/\(\sqrt Hz\), gives margin at the narrowest bandpass settings. Electro-optic sampling in thin GaP, probed by a second pickoff, records the field and hence the spectrum, linewidth and polarisation. The mismatch between laser and beam repetition rates gives 499 laser-only events per coincidence, so laser-generated terahertz from the target is measured and subtracted continuously. Dark current from the RF structure needs a fourth acquisition arm, RF on with the cathode laser blocked, since it reaches the target whenever the RF is on.

\section*{S4. Separation from stationary emission}

The wafer, once conducting, emits diffraction radiation, and it becomes conducting exactly when the switch fires, so timing alone cannot identify the process. The separation is spectral. A boundary that does not change in time can redistribute the frequencies presented to it but cannot create new ones, so coherent diffraction radiation is suppressed twice,

\[S_{DR}(\omega) \propto N^{2}\underset{aperture}{\underbrace{\exp\left( - \frac{2\omega b}{\gamma c} \right)}}\underset{bunching}{\underbrace{\exp\left( - \omega^{2}\sigma_{t}^{2} \right)}},\]

by the finite radial extent of the field\cite{r2,r3} and by the bunch length. At 3 THz, \(b = 2mm\) and \(\sigma_{t} = 150fs\) these are \(1.6 \times 10^{- 7}\) and \(3.4 \times 10^{- 4}\), a combined \(5.3 \times 10^{- 11}\). The aperture factor is taken from the full form \(x^{2}K_{1}^{2}(x)\), \(x = \omega b/\gamma v\), since the exponential asymptote drops a prefactor \(\pi x/2 = 14\) and would understate the background.

The signal carries neither factor. The plasma is not driven at 3 THz; it is handed the instantaneous field and oscillates at a frequency set by the laser, with every electron contributing in phase. Incoherent emission scales as \(N\) with no bunching factor, and equals the coherent background at

\[f_{\times} = \frac{\sqrt{\ln N}}{2\pi\sigma_{t}} \simeq 4.7THz,\]

so the analysis channel is kept below about 4 THz. Figure~S1 shows the separation.

\begin{figure}[t]\centering\includegraphics[width=\linewidth]{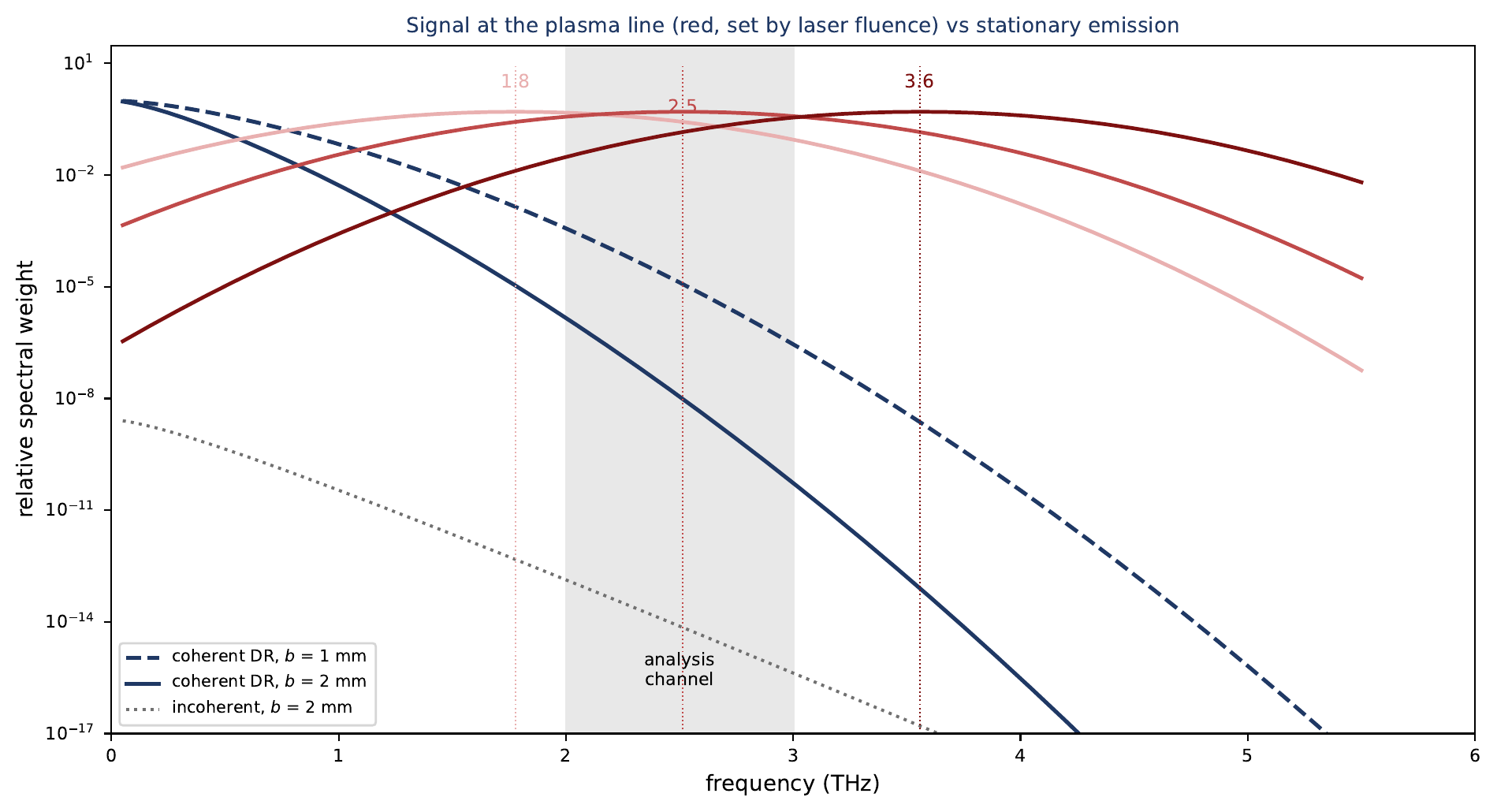}
\caption{Spectral separation. Stationary emission (blue, grey) is confined to the spectrum the bunch delivers and falls exponentially above \(f_{c}\). The temporal-boundary signal (red) is the collective line of the 10 \(\mu m\) membrane at \(0.78\,\omega_{p0}\), fixed by the fluence, in a channel where stationary contributions are suppressed by ten orders of magnitude at \(b = 2mm\). Labels give the line position in THz at 50, 100 and 200 \(\mu J/cm^{2}\).}\label{fig:S1}\end{figure}

\section*{S5. Excited depth and the choice of target}

Deeper layers store more energy, \(U \propto d\), and oscillate longer, since the radiative width falls as \(1/d\); below \(d_{\rm crit}=c(1+n_r)/2(\varepsilon_b-1)\omega_p\), the critical depth of Sec.~2.4, they do not oscillate at all. Two things cap the gain. Collisions hold 0.65 THz of width whatever the depth, so \(Q\) saturates at \(f/\Delta f_{coll} = 5.5\). And the depths must radiate in phase, which requires \(d < \lambda_{m}/2\), 12 \(\mu m\) at 3.6 THz. A 10 \(\mu m\) layer sits at \(0.8\) of that limit with \(Q = 1.0\) to \(2.4\) across the fluence range, and at 800 nm it is also the absorption depth, so the whole layer is excited as one and the grading costs nothing in width. Deeper membranes would not be excited as one and their collective mode would split.

The membrane beats the thick wafer in three ways at the same aperture, material and fluence, two of which multiply into the yield. The resonator is unloaded, gaining \(\left( 1 + n_{r} \right)/2 = 2.2\) in \(Q_{rad}\), which lengthens the oscillation and lowers f\_rad from 0.79 to 0.65, so it buys the spectroscopy rather than yield. The field enters more fully, \(\varepsilon_{eff} = 7.36\) against \(2.78\), because the film is thin against the pulse and the reflections from its two faces partly cancel (Sec.~2.2). And both faces are vacuum, so the forward share is one half rather than \(0.226\). Together: 1.1 nJ detected against 207 pJ. Measuring both tests the field entry directly. With the bunch factors divided out, the ratio of stored energies is the ratio of \(\varepsilon_{eff}\), \(7.36/2.78 = 2.6\), thin-film entry against Fresnel entry; with them in, \(42/14.5 = 2.9\). Table~S1 compares the targets reachable with the existing laser.

\noindent{\small\itshape Table S1. Targets compared at \(b = 1mm\). ``Line'' is the collective plasma line as a fraction of the surface plasma frequency with its \(Q\) across the fluence range; ``none'' means the layer is overdamped and emits a single-cycle transient. Detected energy applies \(f_{rad}\), the forward share and 8\% collection. The last column is the target's own laser-generated terahertz emission at the fluence each option needs.}

\begin{longtable}[]{@{}
  >{\raggedright\arraybackslash}p{(\columnwidth - 12\tabcolsep) * \real{0.1429}}
  >{\raggedright\arraybackslash}p{(\columnwidth - 12\tabcolsep) * \real{0.1429}}
  >{\raggedright\arraybackslash}p{(\columnwidth - 12\tabcolsep) * \real{0.1429}}
  >{\raggedright\arraybackslash}p{(\columnwidth - 12\tabcolsep) * \real{0.1429}}
  >{\raggedright\arraybackslash}p{(\columnwidth - 12\tabcolsep) * \real{0.1429}}
  >{\raggedright\arraybackslash}p{(\columnwidth - 12\tabcolsep) * \real{0.1429}}
  >{\raggedright\arraybackslash}p{(\columnwidth - 12\tabcolsep) * \real{0.1429}}@{}}
\toprule\noalign{}
\begin{minipage}[b]{\linewidth}\raggedright
target, switching
\end{minipage} & \begin{minipage}[b]{\linewidth}\raggedright
\(d\)
\end{minipage} & \begin{minipage}[b]{\linewidth}\raggedright
\(\varepsilon_{eff}\)
\end{minipage} & \begin{minipage}[b]{\linewidth}\raggedright
line, \(Q\)
\end{minipage} & \begin{minipage}[b]{\linewidth}\raggedright
\(U\) (nJ)
\end{minipage} & \begin{minipage}[b]{\linewidth}\raggedright
detected
\end{minipage} & \begin{minipage}[b]{\linewidth}\raggedright
bkg. (pJ)
\end{minipage} \\
\midrule\noalign{}
\endfirsthead
\toprule\noalign{}
\begin{minipage}[b]{\linewidth}\raggedright
target, switching
\end{minipage} & \begin{minipage}[b]{\linewidth}\raggedright
\(d\)
\end{minipage} & \begin{minipage}[b]{\linewidth}\raggedright
\(\varepsilon_{eff}\)
\end{minipage} & \begin{minipage}[b]{\linewidth}\raggedright
line, \(Q\)
\end{minipage} & \begin{minipage}[b]{\linewidth}\raggedright
\(U\) (nJ)
\end{minipage} & \begin{minipage}[b]{\linewidth}\raggedright
detected
\end{minipage} & \begin{minipage}[b]{\linewidth}\raggedright
bkg. (pJ)
\end{minipage} \\
\midrule\noalign{}
\endhead
\bottomrule\noalign{}
\endlastfoot
Si wafer, 800 nm & 10 \(\mu m\), graded & 2.78 & \(0.8\,\omega_{p0}\), \(\sim 1\) & 14.5 & 207 pJ & \(< 2\) \\
Si membrane, 800 nm & 10 \(\mu m\), graded & 7.36 & \(0.78\,\omega_{p0}\), 1.0 to 2.4 & 42 & 1.1 nJ & \(< 2\) \\
Si membrane, 800 nm & 1 \(\mu m\), uniform & 7.36 to 11.7 & none, \(< 1\) & 5.6 to 8.9 & 43 to 280 pJ & \(< 0.3\) \\
GaAs wafer, 800 nm & 1 \(\mu m\), graded & 2.79 & none, \(< 1\) & 2.1 & 7 to 17 pJ & 3.5 to 31 \\
Si wafer, 400 nm & 0.1 \(\mu m\), graded & 2.78 & none, \(< 1\) & 0.22 & \(\sim 0.6pJ\) & \(< 0.1\) \\
\end{longtable}

The membrane is mechanically comfortable. At 200 \(\mu J/cm^{2}\) the fluence is three orders below the ablation threshold; the membrane absorbs 38 per cent of the pump, 38 mW/\(cm^{2}\) at 500 Hz, which a 10 \(\mu m\) sheet conducts to a 3 mm frame with a rise below one kelvin. A span to \(3b\) captures 89 per cent of the stored energy, so a 6 mm free-standing membrane with a 2 mm aperture suffices; 10 \(\mu m\) silicon membranes on windows of that size are catalogue parts. A 6.5 MeV electron crossing 10 \(\mu m\) of silicon loses 4 keV, so halo on the membrane is harmless. Gallium arsenide is listed for comparison only: at 1 \(\mu m\) it is overdamped, it gains none of the membrane's advantages, and its laser-only emission is comparable to its signal (Sec.~S7).

\section*{S6. Limits of the tuning range}

The frequency is set by the carrier density alone and the window in which it can be believed by the beam and the aperture, through \(f_{c} = \gamma c/2\pi b\). The usable range is their intersection. At the low end four things degrade together: the quiet window closes, since 1 THz is only \(1.6f_{c}\) at \(b = 1mm\) and the background there is a few per cent rather than \(10^{- 7}\); \(f_{rad}\) holds at 0.65 across the resonant range and falls only below critical damping, under about 1 THz; \(Q\) passes through unity at the 1.7 THz line of 50 \(\mu J/cm^{2}\), below which there is no line; and the moving-source parameter \(\gamma c/\omega_{p}b\) exceeds unity below 0.65 THz. Low-frequency points belong at the larger apertures. At the high end every figure of merit improves with \(\omega_{p}\) until the phase-coherence ceiling of Sec.~S5, near 4.5 THz for the 10 \(\mu m\) layer. The best-conditioned part of the range is \(2\) to 4 THz.

\section*{S7. Laser-only emission from the target}

Any semiconductor surface illuminated by the switching pulse emits terahertz by optical rectification, surface-field acceleration and the photo-Dember effect. In gallium arsenide the conversion efficiency at 800 nm is \(10^{- 5}\) to \(10^{- 6}\), so a few microjoules of pump give a few to tens of picojoules in the signal band, from the same pulse, comparable to the GaAs signal itself. Silicon, centrosymmetric and indirect, is three orders weaker: under 2 pJ against 1.1 nJ at the membrane's fluence. The background is independent of the switch delay and is measured by the interleaved laser-only shots. It is also linearly polarised along the pump, whereas the signal is radially polarised, so a rotating analyser separates them; that wire-grid check is confirmatory for silicon and would be essential for gallium arsenide.

\section*{S8. The yield table and its scaling}

Table~S2 gives the computed yields at \(F = 100\mu J/cm^{2}\) with \(f_{rad} = 0.79\) from (18). Three features govern the design. The detected energies are tens to hundreds of picojoules, above a pyroelectric detector's single-shot threshold at every aperture. The signal falls only as \(b^{- 1.8}\) while the stationary background falls exponentially, by eight orders across the series, so detector sensitivity buys discrimination and the reference aperture is \(b = 2mm\); the two power laws, \(U \propto b^{- 1.8}\) and \(FWHM = 1.20\, b/\gamma c + 0.17ps\), are themselves observables. And the frequency and the energy are controlled independently, Table~S3, which is the signature.

\noindent{\small\itshape Table S2. Yield against aperture radius for the reference beam, from the two-dimensional solution. \(\eta\) is the bunch-length factor of Sec.~S1; forward and detected energies apply \(f_{rad} = 0.79\), the vacuum-side share \(0.226\) and 8\% collection.}

\begin{longtable}[]{@{}
  >{\raggedright\arraybackslash}p{(\columnwidth - 14\tabcolsep) * \real{0.1250}}
  >{\raggedright\arraybackslash}p{(\columnwidth - 14\tabcolsep) * \real{0.1250}}
  >{\raggedright\arraybackslash}p{(\columnwidth - 14\tabcolsep) * \real{0.1250}}
  >{\raggedright\arraybackslash}p{(\columnwidth - 14\tabcolsep) * \real{0.1250}}
  >{\raggedright\arraybackslash}p{(\columnwidth - 14\tabcolsep) * \real{0.1250}}
  >{\raggedright\arraybackslash}p{(\columnwidth - 14\tabcolsep) * \real{0.1250}}
  >{\raggedright\arraybackslash}p{(\columnwidth - 14\tabcolsep) * \real{0.1250}}
  >{\raggedright\arraybackslash}p{(\columnwidth - 14\tabcolsep) * \real{0.1250}}@{}}
\toprule\noalign{}
\begin{minipage}[b]{\linewidth}\raggedright
\(b\) (mm)
\end{minipage} & \begin{minipage}[b]{\linewidth}\raggedright
\(U\) (nJ)
\end{minipage} & \begin{minipage}[b]{\linewidth}\raggedright
\(U/U_{vac}\)
\end{minipage} & \begin{minipage}[b]{\linewidth}\raggedright
\(\eta\)
\end{minipage} & \begin{minipage}[b]{\linewidth}\raggedright
FWHM (ps)
\end{minipage} & \begin{minipage}[b]{\linewidth}\raggedright
forward (nJ)
\end{minipage} & \begin{minipage}[b]{\linewidth}\raggedright
detected (pJ)
\end{minipage} & \begin{minipage}[b]{\linewidth}\raggedright
DR suppression
\end{minipage} \\
\midrule\noalign{}
\endfirsthead
\toprule\noalign{}
\begin{minipage}[b]{\linewidth}\raggedright
\(b\) (mm)
\end{minipage} & \begin{minipage}[b]{\linewidth}\raggedright
\(U\) (nJ)
\end{minipage} & \begin{minipage}[b]{\linewidth}\raggedright
\(U/U_{vac}\)
\end{minipage} & \begin{minipage}[b]{\linewidth}\raggedright
\(\eta\)
\end{minipage} & \begin{minipage}[b]{\linewidth}\raggedright
FWHM (ps)
\end{minipage} & \begin{minipage}[b]{\linewidth}\raggedright
forward (nJ)
\end{minipage} & \begin{minipage}[b]{\linewidth}\raggedright
detected (pJ)
\end{minipage} & \begin{minipage}[b]{\linewidth}\raggedright
DR suppression
\end{minipage} \\
\midrule\noalign{}
\endhead
\bottomrule\noalign{}
\endlastfoot
0.75 & 22.4 & 0.237 & 0.60 & 0.40 & 4.0 & 319 & \(2.2 \times 10^{- 6}\) \\
1.0 & 14.5 & 0.238 & 0.69 & 0.48 & 2.6 & 207 & \(2.8 \times 10^{- 7}\) \\
1.5 & 7.2 & 0.228 & 0.80 & 0.58 & 1.25 & 103 & \(4.1 \times 10^{- 9}\) \\
2.0 & 4.2 & 0.216 & 0.87 & 0.74 & 0.73 & 59 & \(5.3 \times 10^{- 11}\) \\
3.0 & 1.8 & 0.193 & 0.93 & 1.06 & 0.31 & 25 & \(7.9 \times 10^{- 15}\) \\
\end{longtable}

\begin{figure}[t]\centering\includegraphics[width=\linewidth]{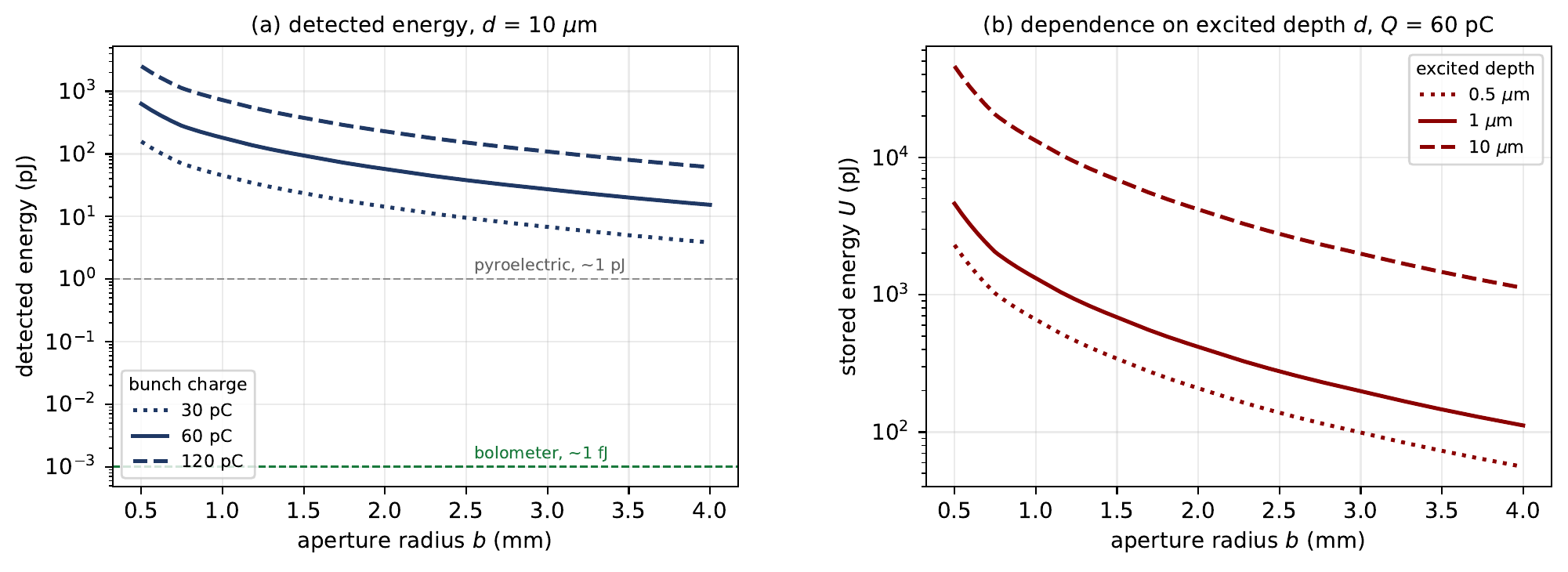}
\caption{(a) Detected energy against aperture radius and bunch charge, with single-shot floors for pyroelectric and bolometric detection. (b) Stored energy against aperture radius for three excited depths at 60 pC, all at the thick-wafer \(\varepsilon_{eff}\).}\label{fig:S2}\end{figure}

\noindent{\small\itshape Table S3. Emission frequency against switching fluence from (19) and (20). Silicon at 800 nm: \(\alpha = 850\,\mathrm{cm}^{-1}\), \(R = 0.33\)\cite{r36}, \(m_{e}^{*} = 0.26\, m_{0}\), \(m_{h}^{*} = 0.37\, m_{0}\)\cite{r35}. Gallium arsenide, for comparison: \(\alpha \simeq\)\(10^{4}\,\mathrm{cm}^{-1}\), \(R = 0.33\), \(m_{e}^{*} = 0.067\, m_{0}\), \(m_{h}^{*} = 0.45\, m_{0}\), \(\varepsilon_{b} = 12.9\). The silicon column is the surface density; the collective line sits near \(0.78\) of the surface plasma frequency.}

\begin{longtable}[]{@{}
  >{\raggedright\arraybackslash}p{(\columnwidth - 10\tabcolsep) * \real{0.1667}}
  >{\raggedright\arraybackslash}p{(\columnwidth - 10\tabcolsep) * \real{0.1667}}
  >{\raggedright\arraybackslash}p{(\columnwidth - 10\tabcolsep) * \real{0.1667}}
  >{\raggedright\arraybackslash}p{(\columnwidth - 10\tabcolsep) * \real{0.1667}}
  >{\raggedright\arraybackslash}p{(\columnwidth - 10\tabcolsep) * \real{0.1667}}
  >{\raggedright\arraybackslash}p{(\columnwidth - 10\tabcolsep) * \real{0.1667}}@{}}
\toprule\noalign{}
\multicolumn{3}{@{}l}{%
\begin{minipage}[b]{\linewidth}\raggedright
Silicon, 800 nm
\end{minipage}} & \multicolumn{3}{l@{}}{%
\begin{minipage}[b]{\linewidth}\raggedright
GaAs, 800 nm (comparison)
\end{minipage}} \\
\midrule\noalign{}
\endfirsthead
\toprule\noalign{}
\multicolumn{3}{@{}l}{%
\begin{minipage}[b]{\linewidth}\raggedright
Silicon, 800 nm
\end{minipage}} & \multicolumn{3}{l@{}}{%
\begin{minipage}[b]{\linewidth}\raggedright
GaAs, 800 nm (comparison)
\end{minipage}} \\
\midrule\noalign{}
\endhead
\bottomrule\noalign{}
\endlastfoot
\cmidrule(lr){1-3}\cmidrule(l){4-6}
\(F\) (\(\mu\)J/cm\(^{2}\)) & \(n\) (cm\(^{- 3}\)) & \(f_{p}\) (THz) & \(F\) (\(\mu\)J/cm\(^{2}\)) & \(n\) (cm\(^{- 3}\)) & \(f_{p}\) (THz) \\
10 & \(2.3 \times 10^{16}\) & 1.02 & & & \\
25 & \(5.8 \times 10^{16}\) & 1.61 & 0.35 & \(9.4 \times 10^{15}\) & 1.01 \\
50 & \(1.2 \times 10^{17}\) & 2.28 & 0.60 & \(1.6 \times 10^{16}\) & 1.32 \\
100 & \(2.3 \times 10^{17}\) & 3.22 & 1.0 & \(2.7 \times 10^{16}\) & 1.70 \\
200 & \(4.6 \times 10^{17}\) & 4.56 & 1.8 & \(4.9 \times 10^{16}\) & 2.28 \\
& & & 3.1 & \(8.4 \times 10^{16}\) & 2.99 \\
\end{longtable}

\section*{S9. Validation of the numerical solution}

Four tests bound the two-dimensional solution. With \(\varepsilon_{b} = 1\) the scattered field is zero to machine precision and the stored energy is \(0.9362\) of the exact value at a radial domain of \(4b\); the deficit is the \(\left( b/R_{\max} \right)^{2}\) truncation of a \(1/R^{2}\) integral, and it matches that prediction at \(4b\), \(8b\) and \(16b\) (\(0.9362\), \(0.9830\), \(0.9948\) against \(0.9375\), \(0.9844\), \(0.9961\)), so the truncation is corrected analytically. Halving the cell from 50 to 12.5 \(\mu m\) changes the result by under \(0.2\) per cent. With the domain scaled with \(b\), \(\varepsilon_{eff}\) is scale invariant, \(2.47\) at \(b = 2mm\) against \(2.48\) at 0.1 mm. Across the aperture series it runs from \(2.3\) at 3 mm to \(2.8\) at 0.75 mm, \(2.78\) at 1 mm and \(2.5\) at 2 mm, against the Fresnel \(2.4\) of (5).

Varying the wafer thickness \(T\) from 0.1 to 2 mm changes the peak energy by under 20 per cent, as it should if the transmitted pulse simply propagates in. That range lies above \(b/\gamma = 73\mu m\); below it the film becomes thin against the pulse and the reflections from its faces cancel, and the uniform wafer mesh cannot resolve it. The 10 \(\mu m\) membrane is reached instead by a second run with the dielectric confined to a finite span, which returns \(\varepsilon_{eff} = 7.36\), converging monotonically through \(7.51\), \(7.38\) and \(7.37\) at 2, 4 and 5 cells across the thickness; a one-dimensional thin-film calculation gives \(6.4\), the geometry adding the rest.

Four results of the two-dimensional solution are used in the main text. First, the entry factor: the computed peak energy is 0.19 to 0.24 of what the vacuum field would store across \(r>b\) for apertures from 3 to 0.75 mm, i.e.\ \(\varepsilon_{\rm eff}=2.3\) to 2.8 against the Fresnel value 2.4 of Eq.~(5), so the plane-wave entry of Sec.~2.2 is accurate to within 17 per cent, the deviation coming from the matching at the rim of the annulus and the trapped wake. Second, the temporal width of \(U(t)\) scales with the aperture radius as \(\mathrm{FWHM}=1.20\,b/\gamma c+0.17\)~ps across the five apertures, the intercept being the bunch length. Third, the peak scales as \(U\propto b^{-1.8}\), softened from the \(b^{-2}\) of Eq.~(6) by the finite bunch duration: for the 150 fs bunch the peak is 0.60 to 0.93 of the point-charge value for \(b=0.75\) to 3 mm (the column \(\eta\) of Table~S2), and the stored energies of Table~1 of the main text include this factor. Fourth, the radiated fraction returned by the one-dimensional solver, 0.83 for the 10~\(\mu\)m layer on a thick wafer and 0.84 to 0.87 for the membrane, includes the plasma's reflection of the part of the pulse still arriving after the switch. The fraction of the stored energy alone that leaves as radiation, the \(f_{rad}\) of Eq.~(18) used in Table~1 of the main text, is 0.79 for the wafer and 0.65 for the membrane.

The one-dimensional solution along the depth, with the field from the two-dimensional run as its initial condition and the Drude response with \(\tau = 0.3ps\), gives the line position and width quoted in the main text, and a radiated energy fraction that also counts the plasma's reflection of the pulse tail: \(f_{rad} = 0.83\) for the thick wafer and \(0.84\) to \(0.87\) for the membrane, the line at \(0.78\, f_{p0}\) with exponent \(0.53\), and a width of 1.5 to 1.8 THz of which 0.90 THz is radiative and 0.65 THz collisional. The two-dimensional membrane run resolves the line position but not yet the width, so the widths quoted are the one-dimensional ones. The fraction of the stored energy that leaves as radiation, which is the f\_rad used in the main text, is 0.79 for the wafer and 0.65 for the membrane (Sec. S10).

\section*{S10. Depth and fluence dependence of the quality factor}

Table S4 gives the radiative quality factor Q\_rad of Eq. (17), the total Q of Eq. (16) and the radiated fraction f\_rad of Eq. (18) of the main text against the excited depth, for a layer on the face of a silicon wafer (1 + n\_r = 4.42) at f\_p = 3.6 THz with \(\tau \) = 0.3 ps. Below critical damping, Q\_rad \textless{} 1/2, the field does not cross zero and Q is not defined; f\_rad there is \(\Gamma /\)(\(\Gamma \) + 1/\(\tau \)) with \(\Gamma \simeq \omega_p^{2}/\gamma_r\) the slow decay rate of the overdamped layer. The one-dimensional switched-Drude solution shows the two regimes directly: a 1 \(\mu m\) membrane decays with no zero crossing at any fluence, and a 10 \(\mu m\) layer crosses zero and oscillates.

\noindent{\small\itshape Table S4. Depth dependence at 3.6 THz, silicon face.}

\begin{longtable}[]{@{}
  >{\raggedright\arraybackslash}p{(\columnwidth - 8\tabcolsep) * \real{0.2000}}
  >{\raggedright\arraybackslash}p{(\columnwidth - 8\tabcolsep) * \real{0.2000}}
  >{\raggedright\arraybackslash}p{(\columnwidth - 8\tabcolsep) * \real{0.2000}}
  >{\raggedright\arraybackslash}p{(\columnwidth - 8\tabcolsep) * \real{0.2000}}
  >{\raggedright\arraybackslash}p{(\columnwidth - 8\tabcolsep) * \real{0.2000}}@{}}
\toprule\noalign{}
\begin{minipage}[b]{\linewidth}\raggedright
d (\(\mu m\))
\end{minipage} & \begin{minipage}[b]{\linewidth}\raggedright
Q\_rad
\end{minipage} & \begin{minipage}[b]{\linewidth}\raggedright
regime
\end{minipage} & \begin{minipage}[b]{\linewidth}\raggedright
Q
\end{minipage} & \begin{minipage}[b]{\linewidth}\raggedright
f\_rad
\end{minipage} \\
\midrule\noalign{}
\endhead
\bottomrule\noalign{}
\endlastfoot
1 & 0.18 & transient & -- & 0.55 \\
2.5 & 0.46 & critical & -- & 0.81 \\
5 & 0.91 & resonant & 0.80 & 0.88 \\
10 & 1.83 & resonant & 1.44 & 0.79 \\
\end{longtable}

Table S5 gives the line computed by the one-dimensional switched-Drude solution for the free-standing 10 \(\mu m\) membrane (1 + n\_r = 2) against the switching fluence; the surface plasma frequency f\_p0 follows from Eqs. (19) and (20). Above 50 \(\mu J/cm^{2}\) the line sits at a fixed fraction 0.78 of f\_p0, the absolute width is constant, and Q grows in proportion to the frequency, as Sec. 2.5 of the main text states. A control run with a uniform layer at the mean density of the graded profile returns the same width at every fluence, 0.62 of the line at 100 \(\mu J/cm^{2}\), which is the evidence that the width is radiative and collisional and not a spread of local plasma frequencies. The radiated fraction f\_rad of Eq. (18), the fraction of the stored energy that leaves as radiation, is 0.65 for the membrane and 0.79 for the wafer face and does not depend on fluence; integrating Eqs. (13) and (14) from the switch with the carriers at rest raises it to 0.73 at the lowest fluence, because the first quarter cycle radiates before the carriers are moving. The solver's own radiated energy, which also counts the plasma's reflection of the part of the pulse still arriving after the switch, runs 0.84 to 0.87 and is not the quantity used in the yield of Sec. 4.

\noindent{\small\itshape Table S5. Fluence dependence for the 10 \(\mu m\) membrane, from the time-domain solution.}

\begin{longtable}[]{@{}
  >{\raggedright\arraybackslash}p{(\columnwidth - 10\tabcolsep) * \real{0.1667}}
  >{\raggedright\arraybackslash}p{(\columnwidth - 10\tabcolsep) * \real{0.1667}}
  >{\raggedright\arraybackslash}p{(\columnwidth - 10\tabcolsep) * \real{0.1667}}
  >{\raggedright\arraybackslash}p{(\columnwidth - 10\tabcolsep) * \real{0.1667}}
  >{\raggedright\arraybackslash}p{(\columnwidth - 10\tabcolsep) * \real{0.1667}}
  >{\raggedright\arraybackslash}p{(\columnwidth - 10\tabcolsep) * \real{0.1667}}@{}}
\toprule\noalign{}
\begin{minipage}[b]{\linewidth}\raggedright
F (\(\mu J/cm^{2}\))
\end{minipage} & \begin{minipage}[b]{\linewidth}\raggedright
f\_p0 (THz)
\end{minipage} & \begin{minipage}[b]{\linewidth}\raggedright
line (THz)
\end{minipage} & \begin{minipage}[b]{\linewidth}\raggedright
line / f\_p0
\end{minipage} & \begin{minipage}[b]{\linewidth}\raggedright
FWHM / line
\end{minipage} & \begin{minipage}[b]{\linewidth}\raggedright
Q
\end{minipage} \\
\midrule\noalign{}
\endhead
\bottomrule\noalign{}
\endlastfoot
25 & 1.61 & 1.11 & 0.69 & 1.29 & 0.8 \\
50 & 2.28 & 1.72 & 0.75 & 1.05 & 1.0 \\
100 & 3.22 & 2.52 & 0.78 & 0.62 & 1.6 \\
200 & 4.56 & 3.57 & 0.78 & 0.42 & 2.4 \\
\end{longtable}

\end{document}